\documentclass[prd,aps,floats,amssymb,preprint,nofootinbib]{revtex4}
\pdfoutput=1
\usepackage{epsfig}
\usepackage{amssymb}
\usepackage{bbm}
\usepackage{color}
\usepackage{ulem}

\definecolor{MyDarkBlue}{rgb}{0.15,0.15,0.45}
\usepackage[linktocpage=true]{hyperref}
\hypersetup{
colorlinks=true,
citecolor=MyDarkBlue,
linkcolor=MyDarkBlue,
urlcolor=MyDarkBlue,
pdfauthor={Garrett Goon, Kurt Hinterbichler and Mark Trodden},
pdftitle={Galileons on Cosmological Backgrounds},
pdfsubject={hep-th}
}

\usepackage{enumitem}
\usepackage{graphicx}
\usepackage{subfigure}
\usepackage{amsmath}
\usepackage{bbm}
\usepackage{color}

\newcommand{\nn}{\nonumber\\}

\newcommand{\be}{\begin{equation}}
\newcommand{\ee}{\end{equation}}
\newcommand{\bea}{\begin{eqnarray}}
\newcommand{\eea}{\end{eqnarray}}
\newcommand{\beas}{\begin{eqnarray*}}
\newcommand{\eeas}{\end{eqnarray*}}

\def\({\left(}
\def\){\right)}

\newcommand{\half}{\frac{1}{2}}

\newcommand{\picubed}{\langle \pi\Pi\pi\rangle}
\newcommand{\pifourth}{\langle \pi\Pi^{2}\pi\rangle}
\newcommand{\pififth}{\langle \pi\Pi^{3}\pi\rangle}
\newcommand{\PII}{\langle \Pi\rangle}
\newcommand{\PIIsquared}{\langle \Pi^{2}\rangle}
\newcommand{\PIIcubed}{\langle \Pi^{3}\rangle}
\newcommand{\dG}{\langle f'\rangle}
\newcommand{\ppdG}{\langle \pi f'\pi\rangle}
\newcommand{\ppdGdG}{\langle \pi f'^{2}\pi\rangle}
\newcommand{\ppdGdGdG}{\langle \pi f'^{3}\pi\rangle}
\newcommand{\dGdG}{\langle f'{}^{2}\rangle}
\newcommand{\dGdGdG}{\langle f'{}^{3}\rangle}
\newcommand{\PdG}{\langle \Pi f'\rangle}
\newcommand{\PdGdG}{\langle \Pi f'{}^{2}\rangle}

\newcommand{\ppPdG}{\langle \pi\Pi f'\pi\rangle}

\newcommand{\pPdGPp}{\langle \pi\Pi f'\Pi\pi\rangle}

\newcommand{\pdGPdGp}{\langle \Pi f'\Pi f'\pi\rangle}

\begin{document}


\title{Galileons on Cosmological Backgrounds}
\author{Garrett Goon\footnote{ggoon@physics.upenn.edu}, Kurt Hinterbichler\footnote{kurthi@physics.upenn.edu} and Mark Trodden\footnote{trodden@physics.upenn.edu}}

\affiliation{Center for Particle Cosmology, Department of Physics and Astronomy, University of Pennsylvania,
Philadelphia, Pennsylvania 19104, USA}

\date{\today}

\begin{abstract}
We construct four-dimensional effective field theories of a generalized DBI galileon field, the dynamics of which naturally take
place on a Friedmann-Robertson-Walker spacetime.  The theories are invariant
under non-linear symmetry transformations, which can be thought of as being inherited from five-dimensional bulk Killing symmetries via the
probe brane technique through which they are constructed. The resulting model provides a framework 
in which to explore the cosmological role that galileons may play as the universe evolves.
\end{abstract}

\maketitle

\tableofcontents

\setcounter{footnote}{0}

\section{Introduction}

Galileons are four-dimensional higher-derivative field theories originally discovered as brane bending modes in decoupling limits of higher-dimensional
induced-gravity models such as the Dvali-Gabadadze-Porrati (DGP) model \cite{Dvali:2000hr,Luty:2003vm,Nicolis:2004qq}.  Galileons have two key features: their equations of motion are second order (despite the appearance of higher derivatives in the action), and they possess novel non-linear global symmetries.  Generalized and abstracted away from these origins \cite{Nicolis:2008in}, the galileons now describe a class of theories with interesting properties including radiative stability~\cite{Luty:2003vm,Hinterbichler:2010xn,Burrage:2010cu}, the successful implementation of the Vainshtein mechanism, and the presence of self-accelerating vacuum solutions (see Sec. 4.4 of \cite{Clifton:2011jh} for a review).
As ghost free, higher derivative field theories, the galileons have have been applied to inflation, late time acceleration, and a variety of other cosmological applications~\cite{Silva:2009km,Burrage:2010cu,Mizuno:2010ag,Creminelli:2010qf,DeFelice:2010as,Kobayashi:2011pc,Mota:2010bs,Wyman:2011mp,Agarwal:2011mg,RenauxPetel:2011dv,Gao:2011mz,Gao:2011qe,RenauxPetel:2011uk,Qiu:2011cy}. They have been used as alternative theories to inflation \cite{Creminelli:2010ba,Hinterbichler:2011qk,Levasseur:2011mw}, kinetic braiding theories \cite{Deffayet:2010qz,Deffayet:2011gz}, and appear naturally in ghost-free theories of massive gravity (\cite{deRham:2011by,deRham:2010gu,deRham:2010ik,deRham:2011ca}, see \cite{Hinterbichler:2011tt} for a review).  The theory has been extended to the multi-galileon case \cite{Padilla:2010de,Hinterbichler:2010xn,Padilla:2010ir,Padilla:2010tj}, supersymmetrized \cite{Khoury:2011da}, and generalized to $p$-forms \cite{Deffayet:2009mn}.

Recent progress has been made towards covariantizing the galileons and putting them on curved backgrounds.
Naive covariantization of the galileons leads to third order equations of motion, a problem which is solvable by introducing appropriate non-minimal couplings between the galileons and the curvature tensor \cite{Deffayet:2009wt,Deffayet:2009mn,Deffayet:2011gz}. However, this construction destroys the the interesting global symmetries of the flat-space theory (see however \cite{Germani:2011bc}).  

A method exists to put the galileons on a fixed curved background while preserving the global symmetries \cite{Goon:2011qf,Goon:2011uw,Burrage:2011bt}.
The method is based on a geometric interpretation in which the galileon field $\pi(x)$ is interpreted as an embedding function describing the position of a 3-brane living in a Minkowski bulk \cite{deRham:2010eu}.  The galileon terms arise from the small field limit of 4D Lovelock invariants \cite{Lovelock:1971yv} and the 4D boundary terms associated with the 5D Lovelock invariants (the Myers terms \cite{Myers:1987yn,Miskovic:2007mg}).  By extending this geometric construction to allow for an arbitrary 5D bulk geometry and and an arbitrary brane embedding \cite{Goon:2011qf,Goon:2011uw}, it is possible to put the galileons on any embeddable background.  The non-linear shift symmetries for $\pi$ are then inherited from the isometries of the bulk geometry (for a short review on generalizing galileons, see \cite{Trodden:2011xh}).

The purpose of the present paper is to apply the brane construction to cosmological FRW spacetimes, and to identify the non-linear symmetries of the resulting theories (this possibility was commented on in  \cite{Burrage:2011bt}).
In what follows, we construct galileons on an FRW background embedded in a flat 5D bulk, so that the symmetry group will be the 15-dimensional Poincare group of 5D flat space, of which the 6 symmetries of FRW (spatial translations and rotations) will be linearly realized.  

After a short review of the general geometric construction of the galileons, we introduce bulk coordinates that define a foliation of 5D Minkowski space by spatially flat FRW slices, and we present expressions for the galileon Lagrangians and their symmetries.  These are ghost-free higher derivative scalar theories that live on an FRW space with an arbitrary time dependence for the scale factor, and which possess 9 non-linearly realized shift-like symmetries.  In the process, we provide expressions for a general Gaussian-normal embedding, of which FRW in flat space is just one example.  For FRW, we display the shorter minisuperspace Lagrangians, which themselves may be useful in a number of cosmological settings.  Additionally, we discuss the small $\pi$ limits and explore the existence and stability of simple solutions for $\pi$.

\subsection*{Conventions and notation}

We use the mostly plus metric signature convention.  Tensors are symmetrized with unit weight, i.e $T_{(\mu\nu)}=\half \left(T_{\mu\nu}+T_{\nu\mu}\right)$.  Curvature tensors are defined by $\left [\nabla_{\mu},\nabla_{\nu}\right ]V^{\rho}=R^{\rho}{}_{\sigma\mu\nu}V^{\sigma}$ and $R_{\mu\nu}=R^{\rho}{}_{\mu\rho\nu}$, $R=R^\mu_{\ \mu}$.

\section{Review of the brane construction for DBI galileons}

The general geometric construction of galileons living on arbitrary curved backgrounds was derived in \cite{Goon:2011qf,Goon:2011uw} and will be briefly reviewed here.  The procedure begins with a fixed 5D metric $G_{AB}(X)$ and a 3-brane defined by the embedding functions $X^{A}(x)$, $A\in\{0,1,2,3,5\}$ where $x^{\mu}$, $\mu\in\{0,1,2,3\}$ are the coordinates native to the hypersurface.  The induced metric and extrinsic curvature on the brane are
\begin{align}
\bar{g}_{\mu\nu}&=e^{A}_{\mu}e^{B}_{\nu}G_{AB},\nn
K_{\mu\nu}&=e^{A}_{\mu}e^{B}_{\nu}\nabla_{A}n_{B} \ ,
\label{gandKinduced}
\end{align}
where $e^{A}_{\mu}=\frac{\partial X^{A}}{\partial x^{\mu}}$ are the tangent vectors to the brane, $n^{A}$ is the spacelike normal vector to the brane, and $\nabla_{A}$ is the covariant derivative with respect to the 5D metric $G_{AB}$.  

The action on the brane is an action for the embedding variables $X^A(x)$, and is chosen to be a diffeomorphism scalar constructed from $\bar{g}_{\mu\nu}$, its covariant derivative and its curvature tensor, as well as the extrinsic curvature tensor,
\begin{align}
S&=\int d^{4}x\, \sqrt{-\bar g}\,\mathcal{L}(\bar{g}_{\mu\nu},\bar{\nabla}_{\mu},\bar{R}^{\alpha}{}_{\beta\mu\nu}, K_{\mu\nu})\ ,
\label{generallagrangian0}
\end{align}
so that it is invariant under under gauge symmetries which are reparameterizations of the brane coordinates,
\begin{align}
\delta X^{A}&=\xi^{\mu}(x)\partial_{\mu}X^{A} \ .
\label{branereparameter}
\end{align}
Given any bulk Killing vector $K^{A}(X)$ satisfying the bulk Killing equation
\be \label{killingequation}
K^C\partial_C G_{AB}+\partial_AK^CG_{CB}+\partial_BK^CG_{AC}=0 \ ,
\ee
both the induced metric and the extrinsic curvature tensor (\ref{gandKinduced}), and therefore the action (\ref{generallagrangian0}), are invariant under the action of the global symmetry
\be \delta_{K} X^{A}=K^{A}(X).\label{gaugeglobalsym}\ee

We fix the gauge symmetry by choosing
\begin{align}
X^{\mu}(x)=x^{\mu}, \ \ \ X^{5}(x)=\pi(x) \ ,
\label{preferredgauge}
\end{align} 
thereby yielding an action solely for $\pi(x)$,
\begin{align}
S&=\int d^{4}x\, \sqrt{-\bar g}\,\mathcal{L}(\bar{g}_{\mu\nu},\bar{\nabla}_{\mu},\bar{R}^{\alpha}{}_{\beta\mu\nu}, K_{\mu\nu})\Big|_{X^{\mu}=x^{\mu},\,X^{5}=\pi(x)} \ ,
\label{generallagrangian}
\end{align}
which has no remaining gauge symmetry.

However, a global symmetry transformation (\ref{gaugeglobalsym}) will generally ruin the gauge choice (\ref{preferredgauge}) and to re-fix the gauge we must make a compensating coordinate transformation on the brane by using (\ref{branereparameter}) with $\xi^{\mu}=-K^{\mu}$.  Thus the combined transformation
\begin{align}
\delta \pi&=-K^{\mu}(x,\pi)\partial_{\mu}\pi+K^{5}(x,\pi)
\label{gensymmetry}
\end{align}
is a global symmetry of the gauge fixed action (\ref{generallagrangian}).

Aside from their symmetries, the other defining characteristic of galileon field theories is the absence of derivatives higher than second order in the equations of motion.  Generic choices for the Lagrangian in (\ref{generallagrangian}) will not meet this requirement, but the Lovelock terms and the Myers boundary terms will \cite{deRham:2010eu}.  In 4D there are only four such terms:
\begin{align}
\mathcal{L}_{2}&= -\sqrt{-\bar g},\nn
\mathcal{L}_{3}&= \sqrt{-\bar g}K,\nn
\mathcal{L}_{4}&= -\sqrt{-\bar g}\bar{R},\nn
\mathcal{L}_{5}&= \frac{3}{2}\sqrt{-\bar g}\left [-\frac{1}{3}K^{3}+K_{\mu\nu}^{2}K-\frac{2}{3}K_{\mu\nu}^{3}-2\left (\bar{R}_{\mu\nu}-\frac{1}{2}\bar{R}\bar{g}_{\mu\nu}\right )K^{\mu\nu}\right ] \ ,
\label{deflagrangian2to5}
\end{align}
where all contractions of indices are performed using the induced metric $\bar{g}_{\mu\nu}$ and its inverse.

In addition, there exists a zero derivative ``tadpole" term which is not of the form (\ref{generallagrangian0}) but which obeys the same symmetries.  This term can be interpreted as the proper volume between an $X^{5}=\rm{const.}$ surface and the brane position $\pi(x)$,
\begin{align}
S_{1}&=\int d^{4}x\, \int^{\pi(x)}d\pi'\, \sqrt{-\det{G_{AB}(\pi',x)}} \ .
\label{deflagrangiantadpole}
\end{align}
As we show in Appendix \ref{tadpoleappendix}, this term also respects the global symmetries (\ref{gensymmetry}).  

\section{DBI Galileons on a Gaussian normal foliation}

In this section we calculate the Lagrangians (\ref{deflagrangian2to5}) and (\ref{deflagrangiantadpole}) in the general case of a background metric which is in Gaussian normal form.  The FRW galileon will be a special case of this general form, and we will specialize to it in later sections.

The background metric in Gaussian normal form is
\be 
\label{metricform} 
G_{AB}dX^AdX^B=f_{\mu\nu}(x,w)dx^\mu dx^\nu+dw^2 \ .
\ee
Here $X^5=w$ denotes the Gaussian normal transverse coordinate, and $f_{\mu\nu}(x,w)$ is an arbitrary metric on the leaves of the foliation defined by the constant $w$ surfaces.  Recall that in the physical gauge (\ref{preferredgauge}), the transverse coordinate of the brane is set equal to the scalar field, $w(x)=\pi(x)$.  This extends our earlier analysis \cite{Goon:2011qf,Goon:2011uw}, by relaxing the condition that the extrinsic curvature of constant $\pi$ slices be proportional to the induced metric.

\subsection{Induced quantities and other ingredients}

The induced metric is
\begin{align}
\bar{g}_{\mu\nu}&=f_{\mu\nu}+\partial_{\mu}\pi\partial_{\nu}\pi,
\end{align}
and its inverse is
\begin{align}
\bar{g}^{\mu\nu}=f^{\mu\nu}-\gamma^{2}\partial^{\mu}\pi\partial^{\nu}\pi \ ,
\end{align}
where 
\be 
\gamma\equiv 1/\sqrt{1+(\partial\pi)^{2}} \ ,
\ee
and the indices on the derivatives are raised with $f^{\mu\nu}$, the inverse of $f_{\mu\nu}$.  

To calculate the extrinsic curvature we need to find the normal vector $n^{A}$, which satisfies
\begin{align}
n^{A}e^{B}_{\nu}G_{AB}&=0 \ ,\nn
n^{A}n^{B}G_{AB}&=1 \ ,
\end{align}
where $e^{B}_{\nu}=\frac{\partial X^{B}}{\partial x^{\nu}}$ are the tangent vectors to the brane.  Solving these equations in the gauge (\ref{preferredgauge}) yields 
\be n_{A}=\gamma(-\partial_{\mu}\pi,1).\ee

The extrinsic curvature is given by
\begin{align}
K_{\mu\nu}&=e^{A}_{\mu}e^{B}_{\nu}\nabla_{A}n_{B} \ ,
\end{align}
which can be written as $K_{\mu\nu}=e^{B}_{\nu}\partial_\mu n_{B}-e^{A}_{\mu}e^{B}_{\nu}\Gamma^C_{AB} n_{C}.$  

The $\nabla_{A}$ is a covariant derivative of the bulk metric and so the Christoffel $\Gamma^C_{AB}$ must be calculated with $X^{5}=w$.  The replacement $w\to\pi(x)$ is then made at the end of the calculation.  Using the bulk coordinates in the form (\ref{Xmetric}), the non-zero 5D Christoffels, $\Gamma^{A}_{BC}$,  are
\begin{align}
\Gamma^{\lambda}_{\mu\nu}&=\Gamma^{\lambda}_{\mu\nu}(f),\nn
\Gamma^{5}_{\mu\nu}&=-\frac{1}{2}f'_{\mu\nu},\nn
\Gamma^{\mu}_{5\nu}&=\frac{1}{2}f^{\mu\lambda}f'_{\lambda\nu} \ ,
\end{align}
where primes denote derivatives with respect to $\pi$.  Note that on the right-hand side of the first line, the Christoffels of $f_{\mu\nu}$ are to be calculated with the $\pi$ dependence held fixed.  The extrinsic curvature then reads
\begin{align}
K_{\mu\nu}&=-\gamma{\nabla}_{\mu}{\nabla}_{\nu}\pi+\frac{1}{2}\gamma f'_{\mu\nu}+\gamma\partial^{\lambda}\pi\partial_{(\mu }\pi f'_{\nu) \lambda} \ ,
\end{align}
where ${\nabla}_{\mu}$ is the covariant derivative calculated from $f_{\mu\nu}$ at fixed $\pi$.

The only remaining components needed to calculate the Lagrangians (\ref{deflagrangian2to5}) are expressions for the induced curvature, $\bar{R}^{\rho}{}_{\sigma\mu\nu}$, which arise in $\mathcal{L}_{4}$ and $\mathcal{L}_{5}$.  At this point, we will specialize to a flat bulk for which the 5D curvature tensor vanishes, so that the induced curvature tensor can be expressed solely in terms of the extrinsic curvature tensor and induced metric via the Gauss-Codazzi equations,
\be R^{(5)}_{ABCD}e^A_{\ \mu}e^B_{\ \nu}e^C_{\ \rho}e^D_{\ \sigma}= 0=
\bar R_{\mu\nu\rho\sigma}-K_{\mu \rho}K_{\nu \sigma}+K_{\mu \sigma}K_{\nu \rho}\ . \ee
The expressions for ${\cal L}_4$ and ${\cal L}_5$ in (\ref{deflagrangian2to5}) then reduce to
\begin{align}
\mathcal{L}_{4}&=-\sqrt{-\bar g}\left [K^{2}-K_{\mu\nu}^{2}\right ], \\
\mathcal{L}_{5}&=\sqrt{-\bar g}\left [K^{3}-3K_{\mu\nu}^{2}K
+2K_{\mu\nu}^{3}\right ].
\end{align}

These are all the elements necessary for the calculation of the Lagrangians.

\subsection{The Lagrangians}

We now present the explicit forms for the DBI galileon Lagrangians.  In all cases, we use the definition 
$\gamma=1/\sqrt{1+\left (\partial\pi\right )^{2}}$ to replace $(\partial\pi)^2$ in favor of $\gamma$ (recall that indices on the derivatives are raised with $f^{\mu\nu}$).  In addition, we employ a shorthand notation.  We define $\Pi_{\mu\nu}=\nabla_{\mu}\nabla_{\nu}\pi$, where the covariant derivative $\nabla_{\mu}$ is calculated from $f_{\mu\nu}$ at fixed $\pi$.  $f'_{\mu\nu}$ denotes the derivative of $f_{\mu\nu}(x,\pi)$ with respect to $\pi$.  We use angular brackets $\langle\ldots\rangle$ to denote traces of the enclosed product as matrices, with all contractions performed using $f^{\mu\nu}$.  For example, we have
\begin{align}
\dG&=f^{\mu\nu}\partial_{\pi}f_{\mu\nu},\nn
\PdG&=\Pi_{\mu\nu}f^{\nu\lambda}\(\partial_{\pi}f_{\lambda\sigma}\)f^{\sigma\mu},\nn
\langle \Pi^3\rangle&=\Pi_{\mu\nu}f^{\nu\lambda}\Pi_{\lambda\sigma}f^{\sigma\rho}\Pi_{\rho\kappa}f^{\kappa\mu}.
\end{align}
In addition, when $\pi$ appears within a angled bracket, it does so only at both ends, and denotes contraction with $\nabla_\mu\pi$, for example,
\begin{align}
\ppdG&=\nabla_{\mu}\pi\, f^{\mu\nu}\( \partial_{\pi}f_{\nu\lambda}\)f^{\lambda\sigma} \nabla_{\sigma}\pi,\nn
\ppPdG&=\nabla_{\mu}\pi\, f^{\mu\nu}\Pi_{\nu\lambda} f^{\lambda\sigma}\( \partial_{\pi}f_{\sigma\rho}\)f^{\rho\kappa} \nabla_{\kappa}\pi \ .
\end{align}

Employing this notation, the Lagrangians (\ref{deflagrangiantadpole}) and (\ref{deflagrangian2to5})  are calculated to be (no integrations by parts have been made in obtaining these expressions)
\begin{align*}
\mathcal{L}_1&=\int^{\pi(x)} d\pi' \sqrt{-\det f_{\mu\nu}(x,\pi')}, \nn
\mathcal{L}_{2}&=-\sqrt{-f}\frac{1}{\gamma},\nn
\mathcal{L}_3&=\sqrt{- f}\Big[-\PII+\dfrac{1}{2} \dG+\gamma^{2}\left(\picubed+\dfrac{1}{2} \ppdG\right)\Big],\nn
\mathcal{L}_{4}&=\sqrt{-f}\Big[-\frac{1}{2} \ppdG^2 \gamma ^3-\dG \picubed \gamma ^3-2
   \pifourth \gamma ^3+2 \picubed \PII \gamma
   ^3\nn
   &\quad -\frac{1}{2} \dG \ppdG \gamma ^3+\PII \ppdG
   \gamma ^3-\frac{\dG^2 \gamma }{4}-\PII^2 \gamma
   +\frac{\dGdG \gamma }{4}\nn
   &\quad -\PdG \gamma +\dG \PII
   \gamma +\PIIsquared \gamma +\frac{\ppdGdG \gamma }{2}\Big],
   \end{align*}

   \begin{align}
   \mathcal{L}_{5}&=\sqrt{-f}\Big[3 \picubed \PII^2 \gamma ^4+\frac{3}{4} \dG \ppdG^2
   \gamma ^4-\frac{3}{2} \PII \ppdG^2 \gamma ^4+\frac{3}{4}
   \dG^2 \picubed \gamma ^4\nn
   &\quad -\frac{3}{4} \dGdG
   \picubed \gamma ^4+3 \PdG \picubed \gamma ^4+6
   \pififth \gamma ^4+3 \dG \pifourth \gamma ^4\nn
&\quad -3
   \dG \picubed \PII \gamma ^4-6 \pifourth
   \PII \gamma ^4-3 \picubed \PIIsquared \gamma
   ^4+\frac{3}{8} \dG^2 \ppdG \gamma ^4\nn
   &\quad +\frac{3}{2}
   \PII^2 \ppdG \gamma ^4-\frac{3}{8} \dGdG \ppdG
   \gamma ^4+\frac{3}{2} \PdG \ppdG \gamma ^4\nn
   &\quad -\frac{3}{2}
   \dG \PII \ppdG \gamma ^4-\frac{3}{2} \PIIsquared
   \ppdG \gamma ^4-\frac{3}{2} \picubed \ppdGdG \gamma
   ^4\nn
   &\quad -\frac{3}{4} \ppdG \ppdGdG \gamma ^4-3 \pPdGPp \gamma
   ^4+3 \ppdG \ppPdG \gamma ^4\nn
   &\quad +\frac{\dG^3 \gamma
   ^2}{8}-\PII^3 \gamma ^2+\frac{3}{2} \dG \PII^2 \gamma
   ^2-\frac{3}{8} \dG \dGdG \gamma ^2+\frac{\dGdGdG \gamma
   ^2}{4}\nn
   &\quad +\frac{3}{2} \dG \PdG \gamma ^2-\frac{3 \PdGdG
   \gamma ^2}{2}-\frac{3 \pdGPdGp \gamma ^2}{2}-\frac{3}{4}
   \dG^2 \PII \gamma ^2\nn
   &\quad +\frac{3}{4} \dGdG \PII
   \gamma ^2-3 \PdG \PII \gamma ^2-2 \PIIcubed \gamma
   ^2-\frac{3}{2} \dG \PIIsquared \gamma ^2\nn
   &\quad +3 \PII
   \PIIsquared \gamma ^2+3 \ppdG \gamma ^2-\frac{3}{4} \dG
   \ppdGdG \gamma ^2+\frac{3}{2} \PII \ppdGdG \gamma
   ^2+\frac{3 \ppdGdGdG \gamma ^2}{4}\Big]\label{L5} \ .
\end{align}

The only dynamical field present is $\pi$, and it enters the Lagrangians both explicitly, and implicitly through the metric $f_{\mu\nu}(x,\pi)$ and its covariant derivatives.  Despite the complicated higher derivative structure of these Lagrangians, the equations of motion will contain at most second order time derivatives, so that they describe only the $\pi$ degree of freedom.

\subsection{\label{globsym2}Global symmetries}

As mentioned in the introduction, if the bulk metric possesses Killing vectors $K^{A}(X)$, then the induced metric and extrinsic curvature, and hence actions of the form (\ref{generallagrangian}), are invariant under the transformations (\ref{gensymmetry}).

The algebra of Killing vectors of $G_{AB}$ contains a subalgebra consisting of those Killing vectors
for which $K^5=0$.  This is the subalgebra of Killing vectors which are parallel to the foliation of constant $w$ surfaces, which generates the subgroup of isometries which preserve the foliation.  For such a Killing vector, the $\mu5$ components of the Killing equations (\ref{killingequation}) tell us that $K^\mu$ is independent of $w$, and the $\mu\nu$ components of the Killing equations tell us that $K^\mu(x)$ is a Killing vector of $f_{\mu\nu}(x,w)$, for any $w$.  We choose a basis of this subalgebra with elements indexed by ${\cal I}$, 
\be 
K_{\cal I}^A(X)=\begin{cases} K_{\cal I}^\mu(x) & A=\mu \\ 0 & A=5\end{cases} \ .
\ee 

We now extend this basis to a basis for the algebra of all Killing vectors by adding a suitably chosen set of linearly independent Killing vectors with non-vanishing $K^5$.  We index these with $I$, so that $(K_{\cal I},K_I)$ is a basis of the full algebra of Killing vectors.  From the $55$ component of Killing's equation, we see that $K^5$ must be independent of $w$, so we may write $K^5(x)$.  

A generic symmetry transformation takes the form
\begin{align}
\delta_{K}X^{A}&=a^{\cal I}K^{A}_{\cal I}(X)+a^{I}K_{I}(X) \ ,
\end{align}
where $a^{\cal I}$ and $a^I$ are constant parameters.  It induces the gauge preserving shift symmetry (\ref{gensymmetry})
\begin{align}
(\delta_{K}+\delta_{g,\rm{comp}})\pi&=-a^{\cal I}K^{\mu}_{\cal I}(x)\partial_{\mu}\pi+a^{I}K_{I}^{5}(x)-a^{I}K_{I}^{\mu}(x,\pi)\partial_{\mu}\pi \ ,\label{gensymmetry2}
\end{align}
demonstrating that the $K_{\cal I}$ symmetries are linearly realized, whereas the $K_{I}$ symmetries are non-linearly realized, corresponding to the spontaneous breaking of the bulk symmetry algebra down to the subalgebra which preserves the leaves of the foliation.  If the bulk metric (\ref{metricform}) has Killing vectors, the Lagrangians (\ref{L5}) will have the symmetries (\ref{gensymmetry2}).

\section{DBI Galileons on cosmological spaces}

We now specialize to the case where the brane metric is FRW.  We thus need a Gaussian-normal foliation of 5D Minkowski space by FRW slices.

\subsection{Embedding 4D FRW in 5D Minkowski}
We consider the case of a spatially flat FRW 3-brane embedded in 5D Minkowski space.  Starting from the bulk Minkowski metric with coordinates $Y^{A}$
\begin{align}
ds^{2}&=-\left (dY^{0}\right )^{2}+\left (dY^{1}\right )^{2}+\left (dY^{2}\right )^{2}+\left (dY^{3}\right )^{2}+\left (dY^{5}\right )^{2} \ ,
\label{Ycoords}
\end{align}
we make a change to coordinates to $t, x^{i},w$, where $i=1,2,3$ runs over the spatial indices on the brane\footnote{This is the transformation used in \cite{Deruelle:2000ge}, except that we have not imposed a $Z_{2}$ symmetry.},

\begin{align}
Y^{0}&=S(t,w)\left (\frac{x^{2}}{4}+1-\frac{1}{4H^{2}a^{2}}\right )-\frac{1}{2}\int dt\, \frac{\dot H}{H^{3}a},\nn
Y^{i}&=S(t,w)x^{i},\nn
Y^{5}&=S(t,w)\left (\frac{x^{2}}{4}-1-\frac{1}{4H^{2}a^{2}}\right )-\frac{1}{2}\int dt\, \frac{\dot H}{H^{3}a} \ .
\label{Xcoords}
\end{align}
Here, $a(t)$ is an arbitrary function of $t$ which will become the scale factor of the 4D space, and overdots denote derivatives with respect to $t$.  We have defined $x^{2}\equiv x^{i}x^{j}\delta_{ij}$, $H\equiv{\dot a}/{a}$, and
\be
S(t,w)\equiv a-\dot{a}w.
\ee
 The lower limits on the integrals in (\ref{Xcoords}) are arbitrary, and different choices merely shift the embedding.  In the case of power law expansions $a(t)\sim t^{\alpha}$, $\alpha>0$, taking the lower limit to be zero puts the big bang at the origin of the embedding space.  

In these new coordinates, the Minkowski metric reads
\begin{align}
ds^{2}&=-n^{2}(t,w)dt^{2}+S^{2}(t, w)\delta_{ij}dx^{i}dx^{j}+dw^{2} \ , 
\label{Xmetric}
\end{align}
where
\begin{align}
n(t,w)&\equiv1-\frac{\ddot a}{\dot a	}w \ .
\end{align}

\begin{figure} 
   \centering
   \includegraphics[width=4.3in]{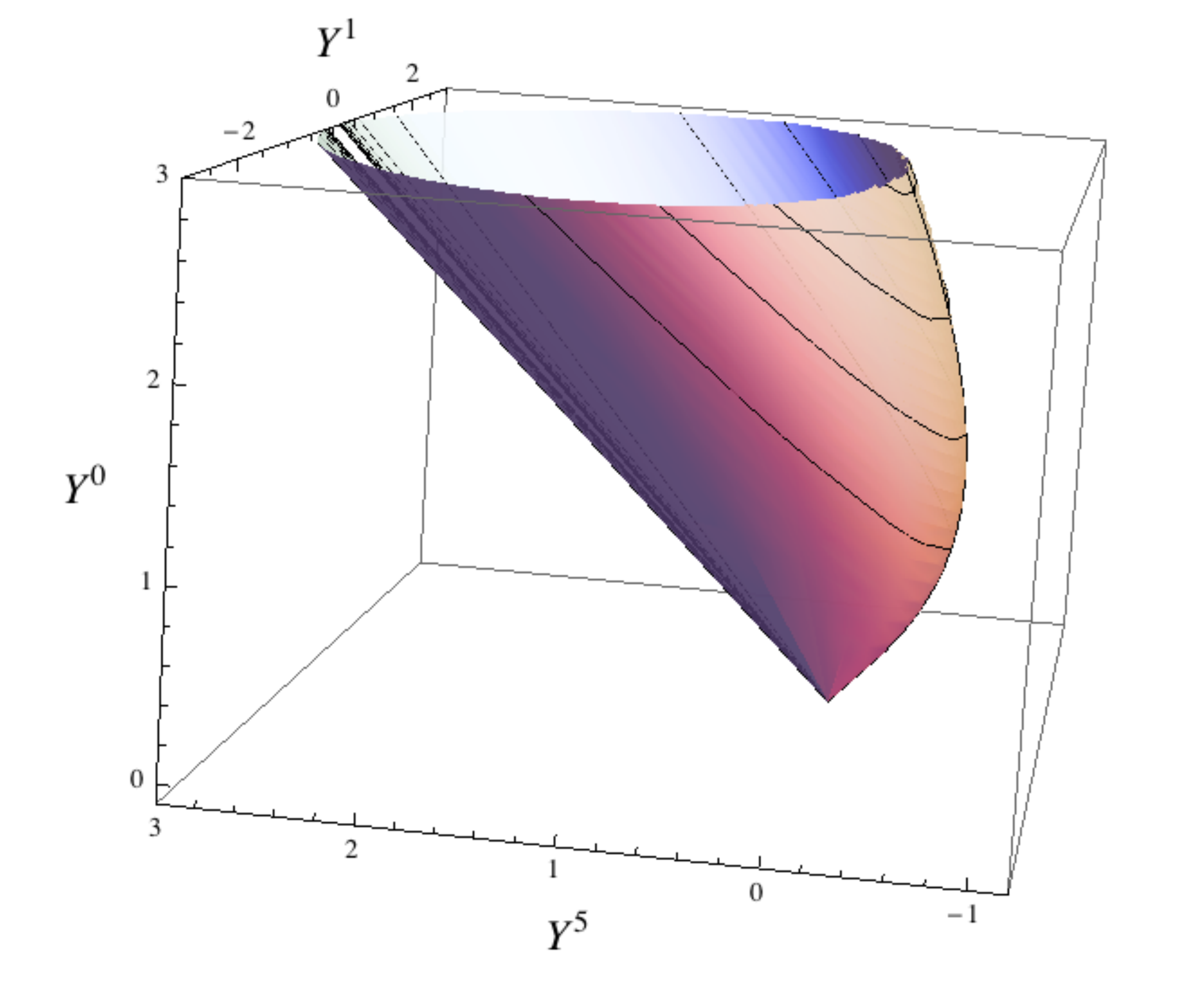}
   \caption{The embedding of an FRW brane in 5D Minksowski space for the case $a(t)=t^{1/2}$.}
   \label{embeddingplot}
\end{figure}

On any $w={\rm const.}$ slice, the induced metric is
\begin{align}
d\tilde{s}^{2}&=-n^{2}(t,w)dt^{2}+S^{2}(t, w)\delta_{ij}dx^{i}dx^{j},
\end{align}
and so after a slice by slice time redefinition $n(t,w)dt=dt'$ we verify that we have indeed foliated $M_{5}$ with spatially flat FRW slices.  Furthermore, the coordinates are Gaussian normal with respect to this foliation. A plot of the embedding in the case $a\sim t^{1/2}$ is shown in Fig.(\ref{embeddingplot}).

In the FRW case, the first two galileon Lagrangians (\ref{L5}) read (no integrations by parts have been made)
\begin{align}
\mathcal{L}_1&=a^3 \pi-\frac{a^2 \Big(3
   \dot{a}^2+a \ddot{a}\Big) \pi ^2}{2 \dot{a}}+a
   \Big(\dot{a}^2+a \ddot{a}\Big) \pi ^3-\frac{1}{4} \dot{a}
   \Big(\dot{a}^2+3 a \ddot{a}\Big) \pi ^4+\frac{1}{5} \ddot{a} \dot{a}^2 \pi ^5,\nn
   \mathcal{L}_{2}&=-(1-\frac{\ddot a}{\dot a}\pi)^{}(a-\dot a\pi)^{3}\sqrt{1-\left(1-\frac{\ddot a}{\dot a}\pi\right)^{-2}\dot\pi^{2}+(a-\dot a\pi)^{-2}(\vec\nabla\pi)^{2}}. \label{explicit12}
   \end{align}   
We relegate the expression for ${\cal L}_3$ to Appendix \ref{L3append}, due to its complexity, and opt not to write out explicit expressions for ${\cal L}_4$ and ${\cal L}_5$ due to their even more unmanageable length. 

\subsection{Global symmetries for FRW}

As reviewed in Section \ref{globsym2}, identifying the relevant global symmetries reduces to the task of finding the Killing vectors of the bulk Minkowski metric in the brane-adapted coordinates (\ref{Xmetric}), 
separating the Killing vectors into those with vanishing $K^{5}$ components, denoted by $K^{A}_{\cal I}$, and those which have non-vanishing $K^{5}$'s, denoted by $K^{A}_{I}$.  

Let $Y^{A}$ be the cartesian coordinates used in (\ref{Ycoords}) with associated basis vectors $\bar{\partial}_{A}$.  The Killing vectors in the $Y^{A}$ coordinates take the form of the ten rotations and boosts, $L_{AB}$, and the five translations $P_{A}$,
\be  L_{AB}=Y_{A}\bar{\partial}_{B}-Y_{B}\bar{\partial}_{A}, \ \ \ \ \ P_{A}=-\bar{\partial}_{A}.
\ee

After rewriting these Killing vectors in terms of the brane-adapted coordinates $\{t,x^i,w\}$ and the associated basis vectors $\{\partial_{t},\partial_i,\partial_w\}$, we find the following combinations which contain no $K^5$ component,
\be
L_{ij}=x^{i}\partial_{j}-x^{j}\partial_{i},\ \ \ \ -\frac{1}{2}\left [L_{i0}+L_{i5}\right] =-\partial_{i} \ .
\ee
These generate the three rotations and three spatial translations of the FRW leaves.  They are the $K^{A}_{\cal I}$.

The remaining vectors form the $K^{A}_{I}$, which we take to be the following combinations,

\begin{align}
v_{i}&=\frac{1}{2}\left [L_{i0}-L_{i5}\right ]= \frac{1}{2} x^{i}\dot{a}  \left [\int dt \, \frac{\dot{H}}{H^3a}\right ] \partial_{w}+\frac{
   x^{i} \big(a-\dot{a}\pi+ \dot{a}^{2}\int dt \, \frac{\dot{H}}{H^3a} 
  \big)}{2\dot{a}-2\pi  \ddot{a}} \partial_{t}\nn
   &\quad -\left [\frac{x^ix^i\dot{a}^{2}+1}{4\dot{a}^{2}} +\frac{
   \int dt \, \frac{\dot{H}}{H^3a}
   }{2a-2\pi 
   \dot{a}}\right ]\partial_{i}+\sum_{j\neq i}\left[-\frac{x^{i}x^{j}}{2}\partial_{j}+\frac{x^jx^j}{4}\partial_{i}\right ] ,\nn
   k_{i}&=-P_{i}=\frac{1}{a-\pi  \dot{a}}\partial_{i}+x^{i} \dot{a}
   \big(\frac{\dot{a} }{\pi 
   \ddot{a}-\dot{a}}\partial_{t}-\partial_{w}\big),\nn
   q&=-\frac{1}{2}\left [P_{0}+P_{5}\right ]=\dot{a} \big(\partial_{w}+\frac{ \dot{a}}{\dot{a}-\pi
    \ddot{a}}\partial_{t}\big),\nn
   u &=-\frac{1}{2}\left [P_{0}-P_{5}\right ]= \frac{x^{2}\dot{a}^{2}-1}{4\dot{a}}\partial_{w}+\frac{
   x^{2} \dot{a}^2+1}{4\dot{a}-4\pi
    \ddot{a}}\partial_{t}-\frac{1
   }{2a-2\pi  \dot{a}}\sum_i x^{i}\partial_{i},\nn
   s&=L_{50} =\left [\frac{a-\pi  \dot{a}+\dot{a}^{2}\int dt \, \frac{\dot{H}}{H^3a}}{\pi\ddot{a}-\dot{a}}\right ]\partial_{t}-\dot{a}\left[\int dt \, \frac{\dot{H}}{H^3a}\right]\partial_{w}+\sum_ix^{i}\partial_{i}\ ,
\end{align}
where $H=\dot a/a$, $x^{2}=\delta_{ij}x^{i}x^{j}$, and the summation convention has been suspended.  The lower limits on the integrals should be the same as those in (\ref{Xcoords}).

The non-linear symmetries of the $\pi$ field are then obtained from (\ref{gensymmetry}), 
\begin{align}
\delta_{v_{i}}\pi&=\frac{1}{2} x^{i}\dot{a}  \int dt \, \frac{\dot{H}}{H^3a} -\frac{
   x^{i} \big(a-\dot{a}\pi+ \dot{a}^{2}\int dt \, \frac{\dot{H}}{H^3a} 
  \big)}{2\dot{a}-2\pi  \ddot{a}} \dot\pi\nn
   &\quad +\left [\frac{x^ix^i\dot{a}^{2}+1}{4\dot{a}^{2}} +\frac{
   \int dt \, \frac{\dot{H}}{H^3a}
   }{2a-2\pi 
   \dot{a}}\right ]\partial_{i}\pi-\sum_{j\neq i}\left[-\frac{x^{i}x^{j}}{2}\partial_{j}\pi+\frac{ x^jx^j}{4}\partial_{i}\pi\right ],\nn
\delta_{k_{i}} \pi&=x^{i} \dot{a} \left(\frac{\dot{a} \dot{\pi}}{\dot{a}-\pi
    \ddot{a}}-1\right)-\frac{\partial_{i}\pi }{a-\pi  \dot{a}},\nn
\delta_q \pi&=\frac{\dot{\pi} \dot{a}^2}{\pi 
   \ddot{a}-\dot{a}}+\dot{a},\nn
 \delta _u \pi &=\frac{x^{2}\dot{a}^{2}-1}{4\dot{a}}-\frac{
   x^{2} \dot{a}^2+1}{4\dot{a}-4\pi
    \ddot{a}}\dot\pi+\frac{1
   }{2a-2\pi  \dot{a}}\sum_i x^{i}\partial_{i}\pi,\nn
  \delta_s   \pi&= -\dot{a}\int dt \, \frac{\dot{H}}{H^3a} +\frac{\left(a-\dot{a}\pi+\dot{a}^{2}
   \int dt \, \frac{\dot{H}}{H^3a}\right) \dot{\pi}}{\dot{a}-\pi  \ddot{a}} -\sum x^{i}\partial_{i}\pi ,  \,\label{fullsyms}
\end{align}
where the replacement $w\to\pi(x^{\mu})$ was performed.

These non-linear transformations are the FRW analogues of the shift symmetries of the flat space galileon.  They are symmetries of the Lagrangians (\ref{explicit12}) and (\ref{L3frw}) as well the ${\cal L}_4$, ${\cal L}_5$ which we did not write out.  Together with the spatial rotation and translation symmetries of FRW, the commutation relations of these transformations are those of the 5D Poincare group.  These are complicated and highly non-linear transformations, and without the brane formalism it would be nearly impossible to guess their form.

\subsection{Minisuperspace Lagrangians}

For cosmological applications where we are not considering fluctuations, we may be most interested in the limiting case in which spatial gradients are set to zero, so that $\pi=\pi(t)$.  In this minisuperspace approximation, the Lagrangians simplify significantly, and we display their full forms here.  In displaying these, the numerators are ordered by increasing powers of $\pi$, and then by patterns of derivatives on the $\pi$ fields.  No integrations by parts have been made.
\begin{align*}
\mathcal{L}_{1}&=a^3 \pi-\frac{a^2 \Big(3
   \dot{a}^2+a \ddot{a}\Big) \pi ^2}{2 \dot{a}}+a
   \Big(\dot{a}^2+a \ddot{a}\Big) \pi ^3-\frac{1}{4} \dot{a}
   \Big(\dot{a}^2+3 a \ddot{a}\Big) \pi ^4+\frac{1}{5} \ddot{a} \dot{a}^2 \pi ^5,\nn
\mathcal{L}_{2}&= -\Big(a-\pi  \dot{a}\Big)^3\sqrt{ \Big(1-\frac{\pi 
   \ddot{a}}{\dot{a}}\Big) ^2-{\dot{\pi}^2}},\nn
\mathcal{L}_{3}&=\Big[3 a^2 \dot{a}^4+a^3 \ddot{a} \dot{a}^2
+\left (-6   a   \dot{a}^5-12   a^2   \ddot{a} 
\dot{a}^3-2   a^3   \ddot{a}^2 \dot{a}\right )\pi
-3     a^2 \dot{a}^4 \dot{\pi}
-   a^3 \dot{a}^3 \ddot{\pi}\nn
&\quad+\left (3  \dot{a}^6+21  a  \ddot{a} 
\dot{a}^4+15  a^2  \ddot{a}^2 \dot{a}^2+ 
a^3  \ddot{a}^3\right )\pi^{2}
+(6    a \dot{a}^5+6  
  a^2  \ddot{a} \dot{a}^3\nn
  &\quad - 
  a^3 \dddot a  \dot{a}^2+ 
  a^3  \ddot{a}^2 \dot{a})\pi\dot\pi+(
-3   a^2  \dot{a}^4-2   a^3  \ddot{a} \dot{a}^2)\dot\pi^{2}+ (3   a^2 \dot{a}^4+  a^3  \ddot{a}  \dot{a}^2)\pi  \ddot{\pi}\nn
&\quad+(-10   \ddot{a} \dot{a}^5-24  a 
\ddot{a}^2 \dot{a}^3-6  a^2  \ddot{a}^3 \dot{a})\pi ^3
+ (-3    \dot{a}^6-12  
  a \ddot{a} \dot{a}^4+3  
  a^2 \dddot a  \dot{a}^3\nn
&\quad-6  
  a^2 \ddot{a}^2 \dot{a}^2)\pi ^2 \dot{\pi}
+(6   a  \dot{a}^5+9   
a^2 \ddot{a} \dot{a}^3)\pi  \dot{\pi}^2+(-3  
 a  \dot{a}^5-3   
a^2  \ddot{a} \dot{a}^3)\pi ^2 \ddot{\pi}\nn
&\quad
+3   a^2 \dot{a}^4 \dot{\pi}^3
+(9  a \dot{a}^2 \ddot{a}^3 +11  \dot{a}^4 
\ddot{a}^2 )\pi^{4}
+(6     \ddot{a} \dot{a}^5-3  
  a  \dddot a  \dot{a}^4+9  
  a  \ddot{a}^2 \dot{a}^3)\pi ^3 \dot{\pi}\nn
&\quad+(
-3    \dot{a}^6-12   a 
 \ddot{a} \dot{a}^4)\pi ^2 \dot{\pi}^2+ 
  (\dot{a}^6+3   a  \ddot{a}  \dot{a}^4)\pi ^3 \ddot{\pi}
-6   a   \dot{a}^5\pi \dot{\pi}^3
-4   \dot{a}^3 \ddot{a}^3\pi ^5\nn
&\quad
+   (\dot{a}^5 \dddot a -4  
  \dot{a}^4  \ddot{a}^2)\pi ^4 \dot{\pi}
+5    \dot{a}^5  \ddot{a}\pi ^3\dot{\pi}^2- 
 \dot{a}^5 \ddot{a} \ddot{\pi}\pi ^4 \nn
&\quad
+3    \dot{a}^6 \dot{\pi}^3\pi ^2
\Big]/\Big[\dot a \left(\left(\dot\pi^2-1\right) \dot a^2+2 \pi  \ddot a
   \dot a-\pi ^2 \ddot a^2\right)\Big],\nn
\end{align*}
\begin{align*}
\mathcal{L}_{4}&=\Big[-6 a \dot{a}^4-6 a^2 \ddot{a} \dot{a}^2
+(6     \dot{a}^5+30   a   \ddot{a} \dot{a}^3+12   a^2   \ddot{a}^2 \dot{a})\pi
+6     a \dot{a}^4 \dot{\pi}
+6    a^2 \dot{a}^3 \ddot{\pi}\nn
&\quad+ (
-24  \ddot{a} \dot{a}^4-42  a  
\ddot{a}^2 \dot{a}^2-6  a^2  \ddot{a}^3)\pi ^2
+(-6     \dot{a}^5-12  
  a  \ddot{a} \dot{a}^3+6  
  a^2  \dddot a  \dot{a}^2\nn
&\quad-6  
  a^2  \ddot{a}^2 \dot{a})\pi  \dot{\pi}
+(6   a  \dot{a}^4+12   a^2  \ddot{a} \dot{a}^2)\dot{\pi}^2+(-12   a  \dot{a}^4-6   a^2   \ddot{a} 
 \dot{a}^2)\pi  \ddot{\pi}\nn
&\quad
+(30   \ddot{a}^2 \dot{a}^3+18  a 
\ddot{a}^3 \dot{a})\pi ^3
+(12     \ddot{a} \dot{a}^4-12  
  a \dddot a  \dot{a}^3+18  
  a  \ddot{a}^2 \dot{a}^2)\pi ^2 \dot{\pi}\nn
&\quad+ (
-6   \dot{a}^5-30   a 
 \ddot{a} \dot{a}^3)\pi  \dot{\pi}^2+(6  
  \dot{a}^5+12   a \ddot{a}  \dot{a}^3)\pi ^2 \ddot{\pi}
-6   a \dot{a}^4 \dot{\pi}^3-12   \dot{a}^2 \ddot{a}^3\pi ^4\nn
&\quad
+(6     \dot{a}^4 \dddot a -12    \dot{a}^3  \ddot{a}^2)\pi ^3 \dot{\pi}
+18    \dot{a}^4  \ddot{a}\pi ^2\dot{\pi}^2-6   \dot{a}^4 \ddot{a} \ddot{\pi}\pi ^3 \nn
&\quad
+6   \dot{a}^5  \pi \dot{\pi}^3\Big]/\Big[  \dot{a}\left(\dot{a} \left(\dot\pi +1\right)-\pi  \ddot{a}\right) \sqrt{ \left(1-{\pi  \ddot{a}\over \dot a}\right)^2 -{\dot\pi ^2}}\Big],
\end{align*}

\begin{align}
\mathcal{L}_{5}&=\Big[-6 \dot{a}^5-18 a \ddot{a} \dot{a}^3
+(36     \ddot{a} \dot{a}^4+36   a   \ddot{a}^2 
\dot{a}^2)\pi
+6     \dot{a}^5 \dot{\pi}\nn
&\quad +18    a \dot{a}^4 \ddot{\pi}+(
-54   \ddot{a}^2 \dot{a}^3-18  a  
\ddot{a}^3 \dot{a})\pi ^2+ (
-12    \ddot{a} \dot{a}^4+18  
  a  \dddot a \dot{a}^3\nn
&\quad -18  
  a  \ddot{a}^2 \dot{a}^2)\pi  \dot{\pi}
+(6    \dot{a}^5+36   a  \ddot{a} 
\dot{a}^3)\dot{\pi}^2+ (-18   \dot{a}^5-18   a   \ddot{a} \dot{a}^3)\pi  \ddot{\pi}
\nn
&\quad +24   \dot{a}^2 \ddot{a}^3\pi ^3
+(24     \dot{a}^3  \ddot{a}^2-18     \dot{a}^4 \dddot a )\pi ^2\dot{\pi}
+18   \dot{a}^4 \ddot{a} \pi ^2 \ddot{\pi}\nn
&\quad -42    \dot{a}^4  \ddot{a}\pi \dot{\pi}^2
-6   \dot{a}^5 \dot{\pi}^3\Big]/\Big[\dot{a} \left(\dot\pi+1\right)-\pi 
   \ddot a\Big]^{2}.\label{minisuperlag}
\end{align}
The $\pi$ equations of motion derived from these are second order in time derivatives.  As before, the scale factor $a(t)$ describes the fixed background cosmological evolution, and does not represent a dynamical degree of freedom. 

Of the symmetries (\ref{fullsyms}), only $\delta _q$ is free of explicit dependence on the spatial coordinates. It is a symmetry of the Lagrangians (\ref{minisuperlag}),
\begin{align}
\delta_q \pi&=\frac{\dot{\pi} \dot{a}^2}{\pi 
   \ddot{a}-\dot{a}}+\dot{a}.\nn
\end{align}

\section{Solutions, fluctuations, and small field limits}

In this section, we explore the existence and stability of simple solutions for $\pi$.  In particular, we focus on the properties of the possible $\pi=0$ solutions.

\subsection{Simple solutions and stability}

Retaining all temporal and spatial derivatives, we expand the Lagrangians to second order in $\pi$, and find, after much integration by parts,

\bea
\mathcal{L}_{1}&=& a^3 \pi-\half  \Big({ \ddot{a}a^3\over \dot a}+3
   \dot{a}a^2\Big) \pi ^2+{\cal O}\(\pi^3\),\nn \\
\mathcal{L}_{2}&=&\(3a^2 \dot a+{a^3\ddot a\over \dot a}\) \pi+\half a^3 \dot\pi^2 -\half a \(\vec\nabla\pi\)^2-3 \left (\ddot{a} a^2+ \dot{a}^2
   a\right )\pi^2+{\cal O}\(\pi^3\), \nn\\
\mathcal{L}_{3}&=& 6 \(a\dot a^2+a^2 \ddot a\)\pi+3 \dot{a} a^2 \dot\pi^2-\left (
   2\dot{a} +\frac{ a \ddot{a}}{\dot{a}} \right )  \(\vec\nabla\pi\)^2-3 \left (3 \dot{a} \ddot{a} a+
   \dot{a}^3\right )\pi^2+{\cal O}\(\pi^3\), \nn\\
\mathcal{L}_{4}&=& 6 \(\dot a ^3+3a \dot a \ddot a\)\pi+9 \dot{a}^2 a\dot\pi^2  -3\left (\frac{\dot{a}^2}{a}+2\ddot a\right )   \(\vec\nabla\pi\)^2 -12 \dot{a}^2 \ddot{a}\pi^2 +{\cal O}\(\pi^3\),\nn\\
\mathcal{L}_{5}&=& 24 \dot a^2 \ddot a\,\pi+12 \dot{a}^3 \dot \pi^2 -12\frac{ \ddot{a}^2 \dot{a}}{a} \(\vec\nabla\pi\)^2+{\cal O}\(\pi^3\).  \nn
\eea
Note that at quadratic order all the higher derivative terms have cancelled out up to total derivative, a consequence of the fact that the equations of motion are second order.

 Consider a theory which is an arbitrary linear combination of the five Lagrangians, 
 \be \mathcal{L}=\sum_{n=1}^5 c_n{\cal L}_n,\ee
 where the $c_n$ are (dimensionful) constants. If $\pi=0$ is to be a solution to the full equations of motion, the linear terms in $\mathcal{L}$ must vanish, which gives the condition
\be c_1 a^3+c_2\(3a^2 \dot a+{a^3\ddot a\over \dot a}\) +6 c_3\(a\dot a^2+a^2 \ddot a\)+6 c_4\(\dot a ^3+3a \dot a \ddot a\)+24 c_5 \dot a^2 \ddot a=0.\label{tadpoleconstraint}\ee
For generic values of the $c_n$, this is a non-linear second order equation for $a(t)$ which can be solved to yield a background for which $\pi=0$ is a solution. If we look for standard power-law solutions, $a(t)=\left (t/t_0\right )^{\alpha}$, the condition (\ref{tadpoleconstraint}) becomes
\be 
\big[24 c_5 (\alpha-1) \alpha^3+ 6 c_4 (4 \alpha-3) \alpha^2 t+6
   c_3 \alpha (2 \alpha-1)t^2+c_2 (4 \alpha-1)t^3+c_1 t^4\big]
   \left(\frac{t}{t_0}\right)^{3 \alpha}=0 \ .
\ee
Each power of $ t$ must vanish independently, so we see that the only non-trivial power-law solutions are for $\alpha=1,3/4,1/2,1/4$.  For these solutions, the corresponding $c_n$ must be non-zero and the others must be set to zero. 

To test the stability around a given solution, we look at the quadratic part of the Lagrangian, which has the following form,
\begin{align}
\mathcal{L}&=\frac{1}{2}A(a(t),c_{n})\dot\pi^{2}-\frac{1}{2}B(a(t),c_{n})(\vec{\nabla}\pi)^{2}-\frac{1}{2}C(a(t),c_{n})\pi^{2} \ ,
\end{align}
where 
\begin{align}
A(a(t),c_{n})&=   c_2 a^3+6  c_3 \dot{a} a^2+18  c_4
   \dot{a}^2 a+24  c_5 \dot{a}^3  , \nn
B(a(t),c_{n})&=   c_2 a+2  c_3\left (
   2\dot{a} +\frac{ a \ddot{a}}{\dot{a}} \right )+  6  c_4 \left (\frac{\dot{a}^2}{a}+2\ddot a\right )+24  c_5\frac{ \ddot{a}
   \dot{a}}{a}
   \ddot{a}, \nn
C(a(t),c_n)&=  c_1\left ( \frac{  \ddot{a} a^3}{\dot{a}}+3   \dot{a}
   a^2\right )+6  c_2 \left (\ddot{a} a^2+ \dot{a}^2
   a\right )+6c_3\left (3 \dot{a} \ddot{a} a+
   \dot{a}^3\right )+24  c_4 \dot{a}^2 \ddot{a}\ .\label{coeffgen}
\end{align}

The stability of the theory against ghost and gradient instability, which is catastrophic at the shortest length scales, requires $A>0$ and $B\ge 0$.   Freedom from tachyon-like instabilities requires $ C\ge 0$.   However a tachyonic instability where $ C<0 $ only affects the large-scale stability of the field, and may be tolerable as long as the time scale associated with the tachyonic mass is of the same order or larger than the Hubble time. 
The equations of motion take the form of a damped harmonic oscillator, $A\ddot{\pi}+\dot{A}\dot{\pi}-B\nabla^{2}\pi+C\pi=0.$
Thus, the time scale $\tau $ associated with a tachyonic mass term is given by $ \tau=\sqrt{A/|C|} $ and the tachyonic instability is tolerable if ${H\tau}\gtrsim 1 $.

\begin{table}\large
\begin{tabular}{|c|c|c|c|c|c|c|c|c|c|}\hline
$\alpha$ & $c_{1}$  & $c_{2}$  & $c_{3}$  & $c_{4}$  & $c_{5}$  & $A$ & $B$ & $C$ & ${H\tau}$\\ \hline \hline
$ 1 $ &  $ 0 $ &  $ 0 $ &  $ 0 $ &  $ 0 $ &  $ c_5 $ &  $ 24 \frac{c_5}{t_0^{3}} $ & $  0  $& $  0  $ & $ 0 $\\ \hline
$ \frac{3}{4} $ &  $ 0 $ &  $ 0 $ &  $ 0 $ &  $ c_4 $ &  $ 0 $ &  $ \frac{81}{8t^{2}}c_{4}\left (t/t_0\right )^{9/4}$& $  \frac{9}{8t^{2}}c_4\, \left (t/t_0\right )^{3/4} $& $  -\frac{81}{32t^{4}}c_4\, \left (t/t_0\right )^{9/4}  $& $  3/2$\\ \hline
$ \frac{1}{2} $ &  $ 0 $ &  $ 0 $ &  $ c_3 $ &  $ 0 $ &  $ 0 $ &  $ \frac{3}{t}c_{3}\left (t/t_0\right )^{3/2} $& $  \frac{1}{t}c_{3}\left (t/t_0\right )^{1/2}  $& $  -\frac{3}{2t^{3}} c_3\, \left (t/t_0\right )^{3/2} $ & ${1\over \sqrt{2}} $\\ \hline
$ \frac{1}{4} $ &  $ 0 $ &  $ c_{2} $ &  $ 0 $ &  $ 0 $ &  $ 0 $ &  $ c_{2}\left (t/t_0\right )^{3/4} $& $  c_2\left (t/t_0\right )^{1/4}  $& $  -\frac{3}{4t^{2}} c_{2} \left (t/t_0\right )^{3/4} $ & ${1\over 2\sqrt{3}} $\\ \hline 
\end{tabular}\caption{Lagrangian coefficients, stability coefficients, and time scale comparisons for fluctuations about $\pi=0$ for all possible non-trivial power law solutions $a(t)=\left (t/t_0\right )^{n}$ .}\label{stabletable}
\end{table}

In Table \ref{stabletable}, we display the coefficients (\ref{coeffgen}) for the four possible power-law solutions.  For the solution $a(t)\sim t$, the choice $c_5>0$ leads to a stable solution, albeit marginally so, since there is no mass or gradient energy.  For each of the other three cases, choosing the relevant coefficient to be positive ensures that $A>0 $, $B>0$, at which point we necessarily have $C<0$ and hence a tachyonic instability.  The tachyon time scale is however $ \tau H\sim  1$ (and happens to be independent of time).   Therefore, each of the four power law solutions are stable to fluctuations over time scales shorter than the age of the universe.

Repeating the analysis in the case of a de-Sitter universe, the condition (\ref{tadpoleconstraint}) for a $\pi=0$ solution becomes
\be c_1+4Hc_2+12 c_3 H^2+24
    c_4 H^3+24  c_5 H^4=0,\ee
and the coefficients (\ref{coeffgen}) of the quadratic part are
\begin{align}
A(a(t),c_{i})&=   a_0^3 e^{3 H t}\( c_2+6 c_3 H+18
    c_4 H^2+24  c_5 H^3\),  \nn
B(a(t),c_{i})&=   a_0 e^{H t}\( c_2+6 c_3 H+18
    c_4 H^2+24  c_5 H^3\), \nn
C(a(t),c_{i})&=  -4 a_0^3 e^{3 H t} H^2 \( c_2+6 c_3 H+18
    c_4 H^2+24  c_5 H^3\).
\end{align}
All the coefficients share a common factor, so the field is either a ghost or a tachyon, in agreement with the findings in Section V.A of \cite{Goon:2011qf}.  Comparing the tachyon time scale against $ 1/H $ gives $ {H\tau}=1/2 $, so the tachyon time scale is approximately the Hubble time.  This would be disastrous for inflation, since the instability would manifest itself after one e-fold, but it may be tolerable for late-time cosmic acceleration.

\subsection{Small $\pi$ symmetries}

The small $\pi$ limits of the symmetries (\ref{fullsyms}) expanded to lowest order in $\pi$, are
\begin{align}
\delta_{v_i}\pi&= \frac{1}{2} x^{i}
\int dt\, \frac{\dot H}{H^{3}a}
\dot{a},\nn
\delta_{k_{i}}\pi&=-x^{i} \dot{a} ,\nn
\delta_{q}\pi&= \dot{a} ,\nn
\delta_{u}\pi&=\frac{
x^2 \dot{a}^2-1}{4 \dot{a}} ,\nn
\delta_{s}\pi&= -
\dot{a}\int dt\, \frac{\dot H}{H^{3}a} \ .
\end{align}

In the case where $\pi=0$ is a solution, these are symmetries of the quadratic action for $\pi$. Otherwise, they are symmetries of the action linear in $\pi$.

\subsection{Galileon-like limits}

When we generate galileon theories by foliating a maximally symmetric bulk by  maximally symmetric branes, as in \cite{Goon:2011uw}, there exist small field limits which greatly simplify the Lagrangians (\ref{L5}).  To take these limits, we form linear combinations $\bar{\mathcal{L}}_{n}=\sum_{m=1}^n c_{n,m}\mathcal{L}_{m}$ of the original Lagrangians, with constant coefficients $c_{n,m}$ chosen such that a perturbative expansion of ${\cal L}_n$ around a constant background $\pi\to\pi_0+\delta{\pi}$ begins at $\mathcal{O}(\delta\pi^{n})$.  In particular, as first shown in \cite{deRham:2010eu}, when applied to the case of a flat brane in a flat bulk, this procedure reproduces the flat space galileons of \cite{Nicolis:2008in}.

The ability to carry out such an expansion appears to be an artifact of maximal symmetry.  The small $\pi$ limit in the present case of a flat bulk and an FRW brane does not, for general $a(t)$, admit a choice of $c_{n,m}$ with the above mentioned properties. 

One case which does work is $a(t)\sim e^{Ht}$, corresponding to a de Sitter brane, which has maximal symmetry.  The induced metric on any $w={\rm const}$ hypersurface is
\begin{align}
ds^{2}&=(1-Hw)^{2}\left [-dt^{2}+e^{2Ht}d\vec{x}^{2}\right]\nn
&=(1-Hw)^{2}\, g_{\mu\nu}^{\,(dS)}dx^{\mu}dx^{\nu} \ ,
\end{align}
where $g_{\mu\nu}^{\,(dS)}$ is the 4D de Sitter metric in inflationary coordinates, and so we are simply foliating 5D minkowski by $dS_{4}$, returning to the setup of a maximally symmetric brane in a maximally symmetric bulk.  In the gauge (\ref{preferredgauge}), the induced metric becomes
\begin{align}
\bar{g}_{\mu\nu}&=\left (-1+H\pi\right )^{2}g_{\mu\nu}^{\,(dS)}+\partial_{\mu}\pi\partial_{\nu}\pi \ .
\label{dsinduced}
\end{align}
If we then make the field redefinition $\tilde{\pi}=-1+H\pi$ and switch to coordinates $\hat{x}^{\mu}=Hx^{\mu}$, the Lagrangians calculated from the induced metric (\ref{dsinduced}) and associated extrinsic curvature take the forms of those in Sec. IV.C of \cite{Goon:2011qf}, from which small $\tilde \pi$ limits can be constructed.

\section{Conclusion}

The probe-brane construction has facilitated the development of entirely new four-dimensional scalar effective field theories with nontrivial symmetries stemming from the Killing symmetries of the higher-dimensional bulk. The simplest example of this construction~\cite{deRham:2010eu} yields flat space galileons~\cite{Nicolis:2008in}, of which the DGP cubic term represents the simplest nontrivial interaction term. In general, however, a much richer structure is possible, depending on the geometries of the bulk and the brane. In previous work~\cite{Goon:2011qf,Goon:2011uw} we have laid out the general framework for deriving new four-dimensional field theories in this way, and have applied the method to the examples in which bulk and brane are maximally symmetric spaces. 

In this paper, we have extended the construction to background geometries with Gaussian normal foliations, of which the cosmological FRW spacetimes are a particularly useful example. We have derived the relevant operators allowed in the Lagrangians, and identified the highly nontrivial symmetry transformations under which they  are invariant. These general expressions are much longer for FRW spacetimes than they are for maximally symmetric ones. By specializing to the minisuperspace approximation, in which the galileons depend only on cosmic time, we are able to provide somewhat more compact versions suitable for understanding the effects of galileons on the background cosmology. However, more complicated questions, such as those involving spatially dependent galileon perturbations, will require the full expressions.  It is possible that integrations by parts would greatly simplify the expressions, but we have not attempted these here.

 We have sought interesting small-field limits of the Lagrangians and their symmetry transformations, as was done for galileons propagating on maximally symmetric backgrounds. Due to the fewer isometries of FRW, the analogous expressions do not seem to exist, except in the special cases in which the FRW space coincides with de Sitter.

Finally, we have studied the stability of simple solutions, namely $\pi=0$ with $ a(t)=\left (t/t_0\right )^{n} $, and find that given a correct sign for coefficients in the Lagrangians, all four possible solutions are stable, at least on the time scales of the background.  One of the four cases leads to a massless field without any gradient energy and the remaining three cases lead to scalar fields with tachyonic masses but the associated time scales are large enough to avoid the potential instability.  For exponential scale factor growth, the $\pi=0$ solution also leads to a tachyon whose time scale is again large enough to stabilize the theory for one e-fold.

\bigskip
\goodbreak
\centerline{\bf Acknowledgements}
\noindent
\\
The authors are grateful to Clare Burrage, Claudia de Rham and Lavinia Heisenberg for helpful conversations. This work is supported in part by NASA ATP grant NNX08AH27G, NSF grant PHY-0930521, and by Department of Energy grant DE-FG05-95ER40893-A020. MT is also supported by the Fay R. and Eugene L. Langberg chair.

\appendix

\section{\label{L3append}Explicit expression for $\mathcal{L}_{3}$}

Here we present the full expression for ${\cal L}_3$ in the FRW case.  No integrations by parts have been made.

   \begin{align*}
\mathcal{L}_{3}&=\Big\{\dot{a}^2 \ddot{a} a^5+3 \dot{a}^4 a^4
-2 \pi  \dot{a} \ddot{a}^2 a^5-14 \pi  \dot{a}^3 \ddot{a} a^4-12 
\pi  \dot{a}^5 a^3
-3 \dot{\pi} a^4 \dot{a}^4
+(\nabla^2\pi) a^3 \dot{a}^3\nn
&\quad -\ddot{\pi} a^5 \dot{a}^3
+18 \pi ^2 a^2 \dot{a}^6+46 \pi ^2 a^3 \ddot{a} \dot{a}^4+19 \pi ^2 
a^4 \ddot{a}^2 \dot{a}^2+\pi ^2 a^5 \ddot{a}^3
+\pi  \dot{\pi} \dot{a} \ddot{a}^2 a^5\nn
&\quad -\pi  \dot{\pi} \dot{a}^2 
a^{(3)} a^5+6 \pi  \dot{\pi} \dot{a}^3 \ddot{a} a^4+12 \pi  \dot\pi \dot{a}^5 a^3
-2 \dot{\pi}^2 \dot{a}^2 \ddot{a} a^5+\pi  \ddot{\pi} \dot{a}^2 \ddot{a} 
a^5-3 \dot{\pi}^2 \dot{a}^4 a^4\nn
&\quad +5 \pi  \ddot{\pi} \dot{a}^4 
a^4+(\nabla\pi)^2  \dot{a}^2 \ddot{a} a^3-3 (\nabla^2\pi) \pi  \dot{a}^2 \ddot{a} a^3
+4 (\nabla\pi)^2  \dot{a}^4 a^2-3 (\nabla^2\pi) 
\pi  \dot{a}^4 a^2\nn
&\quad 
-12 \pi ^3 a \dot{a}^7-64 \pi ^3 a^2 \ddot{a} \dot{a}^5-56 \pi ^3 
a^3 \ddot{a}^2 \dot{a}^3-8 \pi ^3 a^4 \ddot{a}^3 \dot{a}
-18 \pi ^2 \dot{\pi} a^2 \dot{a}^6\nn
&\quad -24 \pi ^2 \dot{\pi} a^3 
\ddot{a} \dot{a}^4+5 \pi ^2 \dot{\pi} a^4 a^{(3)} \dot{a}^3-8 \pi ^2 
\dot{\pi} a^4 \ddot{a}^2 \dot{a}^2
+12 \pi  \dot{\pi}^2 a^3 \dot{a}^5-10 \pi ^2 \ddot{\pi} a^3 
\dot{a}^5
\end{align*}

\vspace{-43pt}

\begin{align*}
&\quad +3 (\nabla^2\pi) \pi ^2 a \dot{a}^5-8 (\nabla\pi)^2  \pi  a \dot{a}^5+13 \pi \dot{\pi}^2 a^4 
\ddot{a} \dot{a}^3-5 \pi ^2 \ddot{\pi} a^4 \ddot{a} \dot{a}^3\nn
&\quad +9 (\nabla^2\pi) 
\pi ^2 a^2 \ddot{a} \dot{a}^3-15 (\nabla\pi)^2  
\pi  a^2 \ddot{a} \dot{a}^3+3 (\nabla^2\pi) \pi ^2 a^3 \ddot{a}^2 \dot{a}
-2 (\nabla\pi)^2  \pi  a^3 \ddot{a}^2 \dot{a}\nn
&\quad 
+3 \dot{\pi}^3 a^4 \dot{a}^4-4 (\nabla\pi)^2  \dot{\pi} 
a^2 \dot{a}^4
-(\nabla^2\pi) \dot{\pi}^2 a^3 \dot{a}^3+2 \nabla\dot\pi\cdot\nabla\pi 
\dot{\pi} a^3 \dot{a}^3\nn
&\quad -(\nabla\pi)^2  \ddot{\pi} a^3 \dot{a}^3+(\nabla\pi)^2  
(\nabla^2\pi) a \dot{a}^3-\delta^{ij}\delta^{kl}\partial_i\pi\partial_j\partial_k\pi\partial_l\pi a \dot{a}^3
+3 \pi ^4 \dot{a}^8\nn
&\quad +41 \pi ^4 a \ddot{a} \dot{a}^6+74 \pi ^4 a^2 
\ddot{a}^2 \dot{a}^4+22 \pi ^4 a^3 \ddot{a}^3 \dot{a}^2
+12 \pi ^3 \dot{\pi} a \dot{a}^7+36 \pi ^3 \dot{\pi} a^2 \ddot{a} 
\dot{a}^5\nn
&\quad -10 \pi ^3 \dot{\pi} a^3 a^{(3)} \dot{a}^4+22 \pi ^3 \dot{\pi} a^3 \ddot{a}^2 \dot{a}^3
-(\nabla^2\pi) \pi ^3 \dot{a}^6+4 (\nabla\pi)^2  \pi ^2 
\dot{a}^6\nn
&\quad -18 \pi ^2 \dot{\pi}^2 a^2 \dot{a}^6+10 \pi ^3 \ddot{\pi} 
a^2 \dot{a}^6-32 \pi ^2 \dot{\pi}^2 a^3 \ddot{a} \dot{a}^4+10 \pi ^3 
\ddot{\pi} a^3 \ddot{a} \dot{a}^4\nn
\end{align*}

\vspace{-66pt}

\begin{align*}
&\quad -9 (\nabla^2\pi) \pi ^3 a 
\ddot{a} \dot{a}^4+27 (\nabla\pi)^2  \pi ^2 a \ddot{a} \dot{a}^4-9 
(\nabla^2\pi) \pi ^3 a^2 \ddot{a}^2 \dot{a}^2\nn
&\quad +18 (\nabla\pi)^2  \pi ^2 a^2 \ddot{a}^2 \dot{a}^2-(\nabla^2\pi) \pi ^3 a^3 \ddot{a}^3+(\nabla\pi)^2  \pi 
^2 a^3 \ddot{a}^3
-12 \pi  \dot{\pi}^3 a^3 \dot{a}^5\nn
&\quad +8 (\nabla\pi)^2  \pi  \dot{\pi}
 a \dot{a}^5+8 (\nabla\pi)^2  \pi  \dot{\pi} a^2 
\ddot{a} \dot{a}^3-(\nabla\pi)^2  \pi  \dot{\pi} a^3 
a^{(3)} \dot{a}^2+(\nabla\pi)^2  \pi  \dot{\pi} a^3 
\ddot{a}^2 \dot{a}\nn
&\quad 
+3 (\nabla^2\pi) \pi  \dot{\pi}^2 a^2 \dot{a}^4-6 \nabla\dot\pi\cdot\nabla\pi 
\pi  \dot{\pi} a^2 \dot{a}^4+3 (\nabla\pi)^2  \pi  \ddot{\pi} a^2 \dot{a}^4
-(\nabla\pi)^2  (\nabla^2\pi) \pi  \dot{a}^4\nn
&\quad +\pi  
\delta^{ij}\delta^{kl}\partial_i\pi\partial_j\partial_k\pi\partial_l\pi \dot{a}^4+(\nabla^2\pi) \pi  \dot{\pi}^2 a^3 
\ddot{a} \dot{a}^2-2 \nabla\dot\pi\cdot\nabla\pi \pi  \dot{\pi} 
a^3 \ddot{a} \dot{a}^2\nn
&\quad +(\nabla\pi)^2  \pi  \ddot{\pi} a^3 
\ddot{a} \dot{a}^2-3 (\nabla\pi)^2  (\nabla^2\pi) 
\pi  a \ddot{a} \dot{a}^2+3 \pi  \delta^{ij}\delta^{kl}\partial_i\pi\partial_j\partial_k\pi\partial_l\pi a
 \ddot{a} \dot{a}^2\nn
&\quad 
-10 \pi ^5 \ddot{a} \dot{a}^7-46 \pi ^5 a \ddot{a}^2 \dot{a}^5-28 \pi ^5
a^2 \ddot{a}^3 \dot{a}^3
-3 \pi ^4 \dot{\pi} \dot{a}^8-24 \pi ^4 \dot{\pi} a \ddot{a} 
\dot{a}^6\nn
\end{align*}

\vspace{-66pt}

\begin{align*}
&\quad +10 \pi ^4 \dot{\pi} a^2 a^{(3)} \dot{a}^5-28 \pi ^4 \dot{\pi}
 a^2 \ddot{a}^2 \dot{a}^4
+12 \pi ^3 \dot{\pi}^2 a \dot{a}^7-5 \pi ^4 \ddot{\pi} a 
\dot{a}^7+3 (\nabla^2\pi) \pi ^4 \ddot{a} \dot{a}^5\nn
&\quad -13 (\nabla\pi)^2 
 \pi ^3 \ddot{a} \dot{a}^5+38 \pi ^3 \dot{\pi}^2 
a^2 \ddot{a} \dot{a}^5-10 \pi ^4 \ddot{\pi} a^2 \ddot{a} \dot{a}^5+9 (\nabla^2\pi)
 \pi ^4 a \ddot{a}^2 \dot{a}^3\nn
&\quad -30 (\nabla\pi)^2  \pi ^3 a \ddot{a}^2 \dot{a}^3
 +3 (\nabla^2\pi) \pi ^4 a^2 \ddot{a}^3 \dot{a}-7 (\nabla\pi)^2  \pi ^3 a^2 \ddot{a}^3 \dot{a}
+18 \pi ^2 \dot{\pi}^3 a^2 \dot{a}^6\nn
&\quad -4 (\nabla\pi)^2  \pi ^2 
\dot{\pi} \dot{a}^6-16 (\nabla\pi)^2  \pi ^2 \dot{\pi} a 
\ddot{a} \dot{a}^4+3 (\nabla\pi)^2  \pi ^2 \dot{\pi} a^2 
a^{(3)} \dot{a}^3-7 (\nabla\pi)^2  \pi ^2 \dot{\pi} a^2 
\ddot{a}^2 \dot{a}^2
\nn
&\quad -3 (\nabla^2\pi) \pi ^2 \dot{\pi}^2 a \dot{a}^5+6
 \nabla\dot\pi\cdot\nabla\pi \pi ^2 \dot{\pi} a \dot{a}^5-3
  (\nabla\pi)^2  \pi ^2 \ddot{\pi} a \dot{a}^5+3 
  (\nabla\pi)^2  (\nabla^2\pi) \pi ^2 \ddot{a} \dot{a}^3\nn
&\quad 
  -3 (\nabla^2\pi) \pi ^2 \dot{\pi}^2 a^2 \ddot{a} \dot{a}^3
  +6 \nabla\dot\pi\cdot\nabla\pi \pi ^2 \dot{\pi} a^2 \ddot{a} 
\dot{a}^3-3 (\nabla\pi)^2  \pi ^2 \ddot{\pi} a^2 \ddot{a} 
\dot{a}^3\nn
\end{align*}

\vspace{-66pt}

\begin{align*}
&\quad -3 \pi ^2 \delta^{ij}\delta^{kl}\partial_i\pi\partial_j\partial_k\pi\partial_l\pi \ddot{a} \dot{a}^3+3 (\nabla\pi)^2 
 (\nabla^2\pi) \pi ^2 a \ddot{a}^2 
\dot{a}\nn
&\quad -3 \pi ^2 \delta^{ij}\delta^{kl}\partial_i\pi\partial_j\partial_k\pi\partial_l\pi a \ddot{a}^2 \dot{a}
+17 a \dot{a}^4 \ddot{a}^3 \pi ^6+11 \dot{a}^6 \ddot{a}^2 \pi ^6
+6 \pi ^5 \dot{\pi} \ddot{a} \dot{a}^7-5 \pi ^5 \dot{\pi} a 
a^{(3)} \dot{a}^6\nn
&\quad +17 \pi ^5 \dot{\pi} a \ddot{a}^2 \dot{a}^5
-3 \pi ^4 \dot{\pi}^2 \dot{a}^8+\pi ^5 \ddot{\pi} \dot{a}^8-22 \pi ^4 
\dot{\pi}^2 a \ddot{a} \dot{a}^6+5 \pi ^5 \ddot{\pi} a \ddot{a} 
\dot{a}^6-3 (\nabla^2\pi) \pi ^5 \ddot{a}^2 \dot{a}^4\nn
&\quad 
+14 (\nabla\pi)^2  \pi ^4 \ddot{a}^2 \dot{a}^4-3 (\nabla^2\pi) 
\pi ^5 a \ddot{a}^3 \dot{a}^2+11 (\nabla\pi)^2  \pi ^4 a 
\ddot{a}^3 \dot{a}^2
-12 \pi ^3 \dot{\pi}^3 a \dot{a}^7\nn
&\quad +8 (\nabla\pi)^2  \pi ^3 \dot{\pi} \ddot{a} \dot{a}^5
-3 (\nabla\pi)^2  \pi ^3 \dot{\pi} 
a a^{(3)} \dot{a}^4+11 (\nabla\pi)^2  \pi ^3 \dot{\pi} 
a \ddot{a}^2 \dot{a}^3
+(\nabla^2\pi) \pi ^3 \dot{\pi}^2 \dot{a}^6\nn
&\quad -2 \nabla\dot\pi\cdot\nabla\pi 
\pi ^3 \dot{\pi} \dot{a}^6+(\nabla\pi)^2  \pi ^3 
\ddot{\pi} \dot{a}^6+3 (\nabla^2\pi) \pi ^3
 \dot{\pi}^2 a \ddot{a} \dot{a}^4-6 \nabla\dot\pi\cdot\nabla\pi 
 \pi ^3 \dot{\pi} a \ddot{a} 
\dot{a}^4\nn 
\end{align*}

\vspace{-66pt}

\begin{align}
&\quad +3 (\nabla\pi)^2  \pi ^3 \ddot{\pi} a \ddot{a} 
\dot{a}^4-3 (\nabla\pi)^2  (\nabla^2\pi) \pi ^3 \ddot{a}^2 
\dot{a}^2+3 \pi ^3 \delta^{ij}\delta^{kl}\partial_i\pi\partial_j\partial_k\pi\partial_l\pi \ddot{a}^2 \dot{a}^2\nn
&\quad -(\nabla\pi)^2  (\nabla^2\pi) \pi ^3 a \ddot{a}^3+\pi ^3 
\delta^{ij}\delta^{kl}\partial_i\pi\partial_j\partial_k\pi\partial_l\pi a \ddot{a}^3
-4 \pi ^7 \dot{a}^5 \ddot{a}^3
+\pi ^6 \dot{\pi} \dot{a}^7 a^{(3)}-4 \pi ^6 \dot{\pi} \dot{a}^6 
\ddot{a}^2\nn
&\quad +5 \pi ^5 \dot{\pi}^2 \ddot{a} \dot{a}^7-\pi ^6 \ddot{\pi} \ddot{a} 
\dot{a}^7+(\nabla^2\pi) \pi ^6 \ddot{a}^3 \dot{a}^3-5 (\nabla\pi)^2  \pi ^5 \ddot{a}^3 \dot{a}^3
+3 \pi ^4 \dot{\pi}^3 \dot{a}^8\nn
&\quad +(\nabla\pi)^2  \pi ^4 \dot{\pi} 
a^{(3)} \dot{a}^5-5 (\nabla\pi)^2  \pi ^4 \dot{\pi} \ddot{a}^2 \dot{a}^4
-(\nabla^2\pi) \pi ^4 \dot{\pi}^2 \ddot{a} \dot{a}^5+2 
\nabla\dot\pi\cdot\nabla\pi \pi ^4 \dot{\pi} \ddot{a} \dot{a}^5\nn
&\quad 
-(\nabla\pi)^2  \pi ^4 \ddot{\pi} \ddot{a} \dot{a}^5+(\nabla\pi)^2 
(\nabla^2\pi) \pi ^4 \ddot{a}^3 \dot{a}-\pi 
^4 \delta^{ij}\delta^{kl}\partial_i\pi\partial_j\partial_k\pi\partial_l\pi \ddot{a}^3 \dot{a}\Big\}/\nn
&\quad  \Big\{\dot{a}^3 (a-\dot{a} \pi )^2
   \dot{\pi}^2-\dot{a} (\dot{a}-\ddot{a} \pi )^2
   \Big((a-\dot{a} \pi )^2+(\vec\nabla\pi)^2\Big)\Big\},\label{L3frw}
\end{align}

where $(\vec\nabla\pi)^{2}=\delta^{ij}\partial_{i}\pi\partial_{j}\pi$ and $\vec{\nabla}^{2}\pi=\delta^{ij}\partial_{i}\partial_{j}\pi$.

\section{\label{tadpoleappendix}Invariance of the tadpole term under global symmetries}

Here we show that the tadpole term (\ref{deflagrangiantadpole}) has the global symmetries (\ref{gensymmetry}).
Under the $\pi$ symmetry (\ref{gensymmetry}), the shift of the tadpole term is
\begin{align}
\delta S_{1}&=\int d^{4}x\, \sqrt{-G(x,\pi)}\left [K^{5}(x,\pi)-K^{\mu}(x,\pi)\partial_{\mu}\pi\right ],
\label{tadpolesymmetry}
\end{align}
where $G(x,\pi)\equiv \det G_{AB}(x,\pi)$.
We will show that the integrand of (\ref{tadpolesymmetry}) is a total derivative by showing that its Euler-Lagrange variation vanishes.  Taking a general variation of the right hand side gives
\begin{align}
&\int d^{4}x\, \quad\sqrt{-G(\pi,x)}\Big\{ \frac{1}{2}\,G^{AB}\partial_{\pi}G_{AB}\,\delta \pi\left [K^{5}(x,\pi)-K^{\mu}(x,\pi)\partial_{\mu}\pi\right ]\nn
&\quad +\left [\partial_{\pi}K^{5}(x,\pi)\delta \pi-\partial_{\pi}K^{\mu}(x,\pi)\delta \pi\partial_{\mu}\pi-K^{\mu}(x,\pi)\partial_{\mu}\delta\pi\right ]\Big\}\nn
&=\int d^{4}x\, \sqrt{-G(\pi,x)}\Big\{ \frac{1}{2}\,G^{AB}\partial_{\pi}G_{AB}\left [K^{5}(x,\pi)-K^{\mu}(x,\pi)\partial_{\mu}\pi\right ]\nn
&\quad +\partial_{\pi}K^{5}(x,\pi)-\partial_{\pi}K^{\mu}(x,\pi)\partial_{\mu}\pi+\partial_{\mu}K^{\mu}(x,\pi)\nn
&\quad+\partial_{\pi}K^{\mu}(x,\pi)\partial_{\mu}\pi +\frac{1}{2}K^{\mu}G^{AB}\partial_{\mu}G_{AB}+\frac{1}{2}K^{\mu}G^{AB}\partial_{\pi}G_{AB}\partial_{\mu}\pi\Big\}\delta\pi\nn
&=\int d^{4}x\, \frac{1}{2}\sqrt{-G(\pi,x)}\, \Big\{G^{AB}\partial_{\pi}G_{AB}K^{5}(x,\pi)+K^{\mu}G^{AB}\partial_{\mu}G_{AB}\nn
&\quad+2\partial_{\pi}K^{5}(x,\pi) +2\partial_{\mu}K^{\mu}(x,\pi)\Big\}\delta \pi \ .
\label{GeneralVariationofTadpole}
\end{align}

Contracting the Killing equation~(\ref{killingequation})
with $G^{AB}$ gives
\begin{align}
G^{AB}\partial_{5}G_{AB}K^{5}(x,\pi)+K^{\mu}G^{AB}\partial_{\mu}G_{AB}+2\partial_{5}K^{5}(x,\pi) +2\partial_{\mu}K^{\mu}(x,\pi)&=0 \ , 
\end{align}
and so (\ref{GeneralVariationofTadpole}) vanishes, indicating that (\ref{tadpolesymmetry}) is a total derivative. Thus, the tadpole term has all the symmetries of the Lagrangians (\ref{generallagrangian}).

\bibliographystyle{utphys}
\bibliography{FRWgalileon-7}

\end{document}